\documentclass[twocolumn,amsmath,secnumarabic,amssymb,nobibnotes, aps,pra]{revtex4-1}
\usepackage[T1]{fontenc}
\usepackage[utf8]{inputenc}
\usepackage{graphicx}
\usepackage{longtable}
\usepackage{bm}
\usepackage{notes2bib}

\begin{document}

\title{Electronic and optical excitations of two-dimensional ZrS$_2$ and HfS$_2$ and their heterostructure}%

\author{Ka Wai Lau}
\author{Caterina Cocchi}
\email{caterina.cocchi@physik.hu-berlin.de}
\author{Claudia Draxl}
\email{claudia.draxl@physik.hu-berlin.de}
\affiliation{Physics Department and IRIS Adlershof, Humboldt Universit\"at zu Berlin, 12489 Berlin, Germany}
\date{\today}%
\begin{abstract}
In a first-principles study based on density-functional theory and many-body perturbation theory, we investigate the electronic properties and the optical excitations of ZrS$_2$ and HfS$_2$ monolayers and their van der Waals heterostructure.
Both materials have an indirect quasi-particle band gap, which amounts to about 2.8 eV in ZrS$_2$ and to 2.6 eV in HfS$_2$. 
In both systems the valence-band maximum is at $\Gamma$ and the conduction-band minimum at M.
Spin-orbit coupling induces a splitting of about 100 meV at the $\Gamma$ point in the valence band, while it does not affect the conduction band.
The optical absorption spectra are dominated by excitonic peaks, with binding energies between 0.6 eV and 0.8 eV.
The ZrS$_2$/HfS$_2$ heterobilayer exhibits a peculiar type-I level alignment with a large degree of hybridization between the two monolayers in the valence band, while the conduction bands retain either ZrS$_2$ or HfS$_2$ character, respectively. 
As a consequence, both the electron and the hole components of the first exciton are localized in the ZrS$_2$ monolayer with non-vanishing probability of finding the hole also in the HfS$_2$ sheet.

\end{abstract}
\maketitle
\section{Introduction}
Since the discovery of monolayer MoS$_2$ as a direct band-gap semiconductor absorbing and emitting light in the visible range~\cite{MoS2Gap,MoS2Gap2}, group-VI transition metal dichalcogenides (TMDCs) have been attracting extensive research~\cite{MoS2Gap2, MoS2Syn, MoS2Kphy, MoS2Kpump, MoS2CT, MoS2pn, bern+13nl}, especially in the field of nano- and opto-electronics~\cite{MoS2mob,yin+11nano}. 
The unique ability of monolayer materials to be combined into vertically-stacked van der Waals (vdW) heterostructures~\cite{geim-grig13nat,het_eng,kim_vdW_het,thyg172DM} offers opportunities to tune the range of absorbed and emitted radiation accessible to these systems~\cite{with+15natm,calm+16apl}.
For example, heterobilayers formed by MoS$_2$ and WS$_2$ exhibit type-II band alignment~\cite{koms-krash13prb,kosm-ferd13prb,amin+15prb}, as desired for photo-diodes, photo-detectors, and analogous opto-electronic devices~\cite{MoS2pn,MoS2CT,yu+14nl,huo+14afm,chen+15nano,wang+18afm}.

The growing popularity of group-VI TMDCs as a new class of semiconducting materials has also stimulated the study of related compounds.
Among them, group-IV TMDCs  have received particular attention due to their promising opto-electronic characteristics~\cite{ZrS2exp, ZrSe2exp, HfS2FET_1, HfS2FET_2, ZrS2sol, wang+16jmcc}.
ZrS$_2$ and HfS$_2$ monolayers combine a band gap in the visible region with exceptionally large charge-carrier mobilities, exceeding even those of group-VI TMDCs~\cite{4TMDrev,TMDmob}.
These features are mainly related to their structural properties.
Different from their group-VI analogs, group-IV TMDCs preferentially crystallize in the so-called 1T arrangement, with octahedral coordination of the metal atom with respect to the chalcogen~\cite{zhua-henn13jpcc}, and remain indirect band-gap semiconductors also in the monolayer form.
While most of these characteristics have been discussed in earlier works~\cite{lv_align,li+14rsc,abdu-joub16pssb,wu+17pe,xin+17pe,zhao+17pssb,latu+18npj2D,rehm-ding18pccp,vu+19sum} based on density-functional theory (DFT), a description of the electronic and optical properties of ZrS$_2$ and HfS$_2$ monolayers that fully accounts for many-body effects is needed to determine their properties beyond the mean-field picture.
This is particularly relevant also to understand and predict the behavior of their heterostructures. 
For example, recent work based on DFT has suggested that the heterobilayer formed by ZrS$_2$ and HfS$_2$ exhibits a type-II level alignment, with the uppermost valence band localized on the HfS$_2$ layer and the lowest conduction band on ZrS$_2$~\cite{rehm-ding18pccp}.

\begin{figure}
\includegraphics[width=0.45\textwidth]{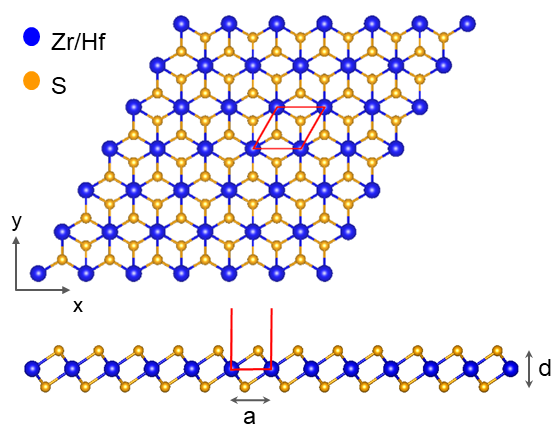}
\caption{Top and side view of the 1T structure of monolayer ZrS$_2$ and HfS$_2$. The unit cell, indicated by the red line, has lattice parameter $a$; $d$ is the distance between the upper and the lower S atoms in the monolayer. Metal atoms are marked in blue and sulfur atoms in orange.}
\label{fig:1T}
\end{figure}

To explore the impact of the electron-electron and the electron-hole interactions in the opto-electronic properties of these materials, we present a first-principles study on monolayer ZrS$_2$ and HfS$_2$ and on their vdW heterostructure based on DFT and many-body perturbation theory (MBPT).
In the framework of DFT we assess the role of spin-orbit coupling (SOC), which is known to induce a sizable energy splitting in the electronic bands of TMDCs~\cite{zhu+11prb,wang+12natn}.
We adopt the $G_0W_0$ approach to obtain the quasi-particle band structures and we solve the Bethe-Salpeter equation to compute optical absorption spectra and excitonic wave-functions.
Our results and their analysis provide information and insight on exciton binding energies, as well as character, composition, and spatial distribution of the electron-hole pairs, thus leading to a deeper understanding of light-matter interaction in these materials.

\section{Theoretical background and computational details}
\label{sec:theory}
Ground-state properties are computed within DFT~\cite{hohe-kohn64pr}.
To effectively simulate monolayer materials, supercells are constructed, including about $30$~\AA{} of vacuum in the non-periodic out-of plane direction to prevent spurious interactions between the replicas. 
All in-plane lattice constants are optimized using the Birch-Murnaghan equation of state~\cite{birch,murnaghan}. 
The Kohn-Sham (KS)~\cite{kohn-sham65pr} eigenvalues and eigenvectors are used as starting point for MBPT calculations. 
The $GW$ approach~\cite{Hedin}, in the single-shot $G_0W_0$ approximation~\cite{hybe-loui85prl}, is adopted to compute the quasi-particle (QP) electronic structure.
The QP energies of the electronic bands $\epsilon_{i\mathbf{k}}$ are given by
\begin{equation}
\epsilon_{i\mathbf{k}}^{QP} = \epsilon_{i\mathbf{k}}^{KS} + Z_{i\mathbf{k}} \left[ \Re \Sigma_{i\mathbf{k}}(\epsilon_{i\mathbf{k}}^{KS}) - V^{xc}_{i\mathbf{k}} \right],
\label{eq:QP}
\end{equation}
where $\Sigma$ is the electronic self-energy, $V^{xc}$ is the exchange-correlation potential, and $Z$ the renormalization factor.
 
The Bethe-Salpeter equation (BSE)~\cite{hank-sham80prb,stri88rnc,rohl-loui00prb} is employed to compute optical absorption spectra accounting for (bound) electron-hole (e-h) pairs.
The BSE is mapped in the eigenvalue problem
\begin{equation}
\sum_{v'c'\mathbf{k'}} \hat{H}^{BSE}_{vc\mathbf{k},v'c'\mathbf{k'}} A^{\lambda}_{v'c'\mathbf{k'}} = E^{\lambda} A^{\lambda}_{vc\mathbf{k}} ,
\label{eq:BSE}
\end{equation}
where the BSE Hamiltonian $\hat{H}^{BSE}_{vc\mathbf{k},v'c'\mathbf{k'}}$ includes the interaction kernel between the electron and the hole.
For further details about the methodology, we refer to Ref.~\cite{FPrevRub}.
In Eq.~\eqref{eq:BSE}, the eigenvalues $E^{\lambda}$ correspond to the excitation energies, and the eigenvectors $A^{\lambda}_{vc\mathbf{k}}$ contain information about the character and composition of the excitations. 
The exciton wave-function is given by the six-dimensional quantity
\begin{equation}
\Psi^{\lambda}(\mathbf{r}_h, \mathbf{r}_e) = \sum_{vc\mathbf{k}} A^{\lambda}_{vc\mathbf{k}} \phi^*_{v\mathbf{k}}(\mathbf{r}_h) \phi_{c\mathbf{k}}(\mathbf{r}_e) ,
\label{eq:exciton}
\end{equation}
where $\phi_{v\mathbf{k}}(\mathbf{r}_h)$ and $\phi_{c\mathbf{k}}(\mathbf{r}_e)$ are the occupied and unoccupied QP states, respectively, that are included in the transition space for the solution of the BSE.
The exciton character and composition in reciprocal space is given by the so-called exciton \textit{weights} 
\begin{equation}
w^{\lambda}_{v\mathbf{k}} = \sum_c |A^{\lambda}_{vc\mathbf{k}}|^2
\label{eq:wh}
\end{equation}
and 
\begin{equation}
w^{\lambda}_{c\mathbf{k}} = \sum_v |A^{\lambda}_{vc\mathbf{k}}|^2,
\label{eq:we}
\end{equation}
which contain information about the eigenvectors (Eq.~\ref{eq:BSE}) and represent the contributions to a given electronic transition to the $\lambda^{th}$ solution of the BSE (see, for example, Refs.~\cite{cocc-drax17jpcm,aggo+17jpcl,turk+19ats}).
Both eigenenergies and eigenvectors of Eq.~\eqref{eq:BSE} enter the expression of the imaginary part of the macroscopic dielectric function
\begin{equation}
\Im\epsilon_M = \dfrac{8\pi^2}{\Omega} \sum_{\lambda} |\mathbf{t}^{\lambda}|^2 \delta(\omega - E^{\lambda}), 
\label{eq:ImeM}
\end{equation}
where the \textit{transition coefficients} $\mathbf{t}^{\lambda}$ are expressed as:
\begin{equation}
\mathbf{t}^{\lambda}= \sum_{vc\mathbf{k}} A^{\lambda}_{vc\mathbf{k}} \frac{\langle v\mathbf{k}|\widehat{\mathbf{p}}|c\mathbf{k}\rangle}{\varepsilon^{QP}_{c\mathbf{k}} - \varepsilon^{QP}_{v\mathbf{k}}} .
\label{eq:t}
\end{equation}
Eq.~\eqref{eq:ImeM} is commonly adopted to represent the optical absorption spectrum and so it is utilized also in the following. 

All calculations presented here are performed with \texttt{exciting}~\cite{exciting}, an all-electron full potential code for DFT and MBPT, implementing the linearized augmented plane-wave plus local orbitals basis set. 
DFT calculations are carried out on a $18\times 18\times 1$ \textbf{k}-mesh. 
In each material we choose muffin-tin (MT) spheres with equal radius for all the atomic species, namely 2.0 bohr in HfS$_2$, 2.2 bohr in ZrS$_2$, and 2.1 bohr in the ZrS$_2$/HfS$_2$ heterostructure. A value of $R_{MT}|\mathbf{G}+\mathbf{k}|_{max} = 8$ is adopted in all systems and all calculations.
This corresponds to a maximum $|\mathbf{G}+\mathbf{k}|$ vector ($|\mathbf{G}+\mathbf{k}|_{max}$) in the interstitial region of 4.0 bohr in HfS$_2$, 3.63 bohr in ZrS$_2$, and 3.81 bohr in the heterostructure~\bibnote{In terms of kinetic energy these values correspond to cutoffs of approximately 215 eV, 175 eV, and 195 eV for HfS$_2$, ZrS$_2$, and the ZrS$_2$/HfS$_2$ heterostructure, respectively.
Note, however, that these quantities do not have the same meaning as in plane-wave codes, since here plane-waves are used only in the interstitial region while a large fraction of the density lies inside the atomic sphere.}.
In all DFT calculations the Perdew-Burke-Ernzerhof (PBE) parameterization~\cite{PBE} of the generalized gradient approximation for the exchange-correlation functional is adopted.
In the heterostructure, the vdW interactions between the monolayers are included by means of Grimme's DFT-D2 functional~\cite{DFTD2}. 
In the DFT calculations, SOC is accounted for within the second-variational formalism~\cite{exciting}. 

$G_0W_0$ calculations~\cite{Exciting2} are run on a $18\times 18\times 1$ \textbf{k}-mesh using 100, 200, and 300 empty states to compute the screened Coulomb interaction within the random-phase approximation in monolayer HfS$_2$, ZrS$_2$, and in the ZrS$_2$/HfS$_2$ heterostructure, respectively.
A truncation of the Coulomb interaction is applied in the $G_0W_0$ calculations, following the scheme proposed by Ismail-Beigi~\cite{CoulCutoff}.
In the BSE calculations~\cite{BSE-exciting} performed within the Tamm-Dancoff approximation, we apply a scissors shift to the conduction bands, with the value taken from the $G_0W_0$ correction of the direct band gap.
In the BSE calculations, a $60\times 60\times 1$ \textbf{k}-mesh shifted from the $\Gamma$ point is adopted for the isolated monolayers.
Three (five) valence bands and two (three) conduction bands are included to describe the transition space in HfS$_2$ (ZrS$_2$), and about 380 \textbf{G+q} vectors ensure an accurate treatment of local-field effects. 
For the heterostructure, a $24\times 24\times 1$ \textbf{k}-mesh is used, $4$ valence and $4$ conduction bands are included, and 395 \textbf{G+q} vectors are taken into account.
We did not apply a truncation of the Coulomb interaction in the BSE calculations where the adopted unit-cell sizes are large enough to yield converged results for neutral excitations. 
Overall, the above-listed computational parameters ensure convergence on the QP energy gaps and on the exciton binding energies within a few tens of meV.

Input and output data are stored in the NOMAD Repository and are freely available for download at the following link: http://dx.doi.org/10.17172/NOMAD/2019.04.08-1.

\section{Results}
\subsection{Monolayer ZrS$_2$ and HfS$_2$} 
\label{mono}
\begin{figure}
\includegraphics[width=0.45\textwidth]{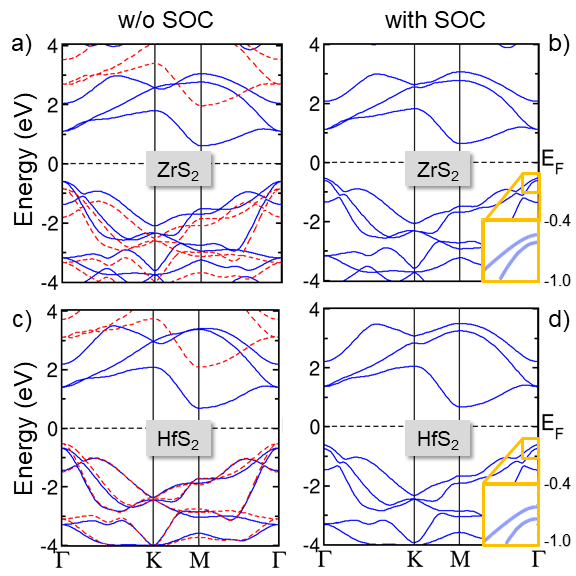}
\caption{Band structures of monolayer ZrS$_2$ (top) and HfS$_2$ (bottom) without SOC (left panels) and with SOC (right panels). Dashed red (solid blue) lines indicate $G_0W_0$ (DFT) results. The Fermi energy ($E_F$) is set to zero in the mid-gap. Inset of panels b) and d): Valence band maximum around $\Gamma$ showing the spin-orbit splitting.}
\label{fig:monobs}
\end{figure}

We start our analysis from the monolayers ZrS$_{2}$ and HfS$_{2}$.
As discussed in the Introduction, the most stable configuration of group-IV TMDCs is the 1T structure~\cite{MoS2_2H-1T_1, MoS2_2H-1T_2,basto+19prm}, where the three atomic layers that form the TMDCs are not vertically aligned on top of each other but horizontally displaced (see Fig.~\ref{fig:1T}).
The resulting crystal structure is similar to a honeycomb lattice but with an extra chalcogen atom located in the center of the hexagons.
Upon structural optimization we obtain in-plane lattice constants ($a$ in Fig.~\ref{fig:1T}) of  3.69 \AA{} and 3.66 \AA{} for ZrS$_{2}$ and HfS$_{2}$, respectively, in agreement with previous DFT calculations~\cite{rasm-thyg15jpcc,zhao+17pssb,rehm-ding18pccp}.
The vertical distance between the two layers of S atoms ($d$ in Fig.~\ref{fig:1T}) is equal to 5.48 \AA{} in ZrS$_2$ and to 5.70 \AA{} in HfS$_2$.

\begin{figure}
\includegraphics[width=0.5\textwidth]{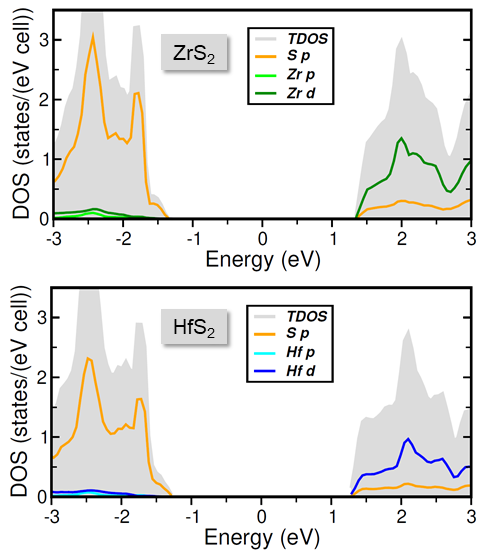}
\caption{Total QP density of states (TDOS, grey area) of monolayer ZrS$_2$ (top panel) and HfS$_2$ (bottom panel). Atom-projected contributions from the dominant states are shown by colored lines according to the legend.}
\label{fig:pdos}
\end{figure}

The band structures of the two monolayers are shown in Fig.~\ref{fig:monobs}, where the results obtained from DFT (solid lines) and $G_0W_0$ (dashed lines) are overlaid on the left panels.
DFT results including SOC are reported in the right panels.
The band structures of ZrS$_2$ and HfS$_2$ are very similar to each other, both exhibiting an indirect band-gap with the valence band maximum (VBM) at $\Gamma$ and the conduction band minimum (CBm) at M.
In ZrS$_2$, the PBE gap is 1.21 eV while in HfS$_2$ it is 1.36 eV, in excellent agreement with previous calculations on the same level of theory~\cite{kang+15pccp,vu+19sum} and with the results reported in the Computational 2D Materials Database (C2DB)~\cite{haas+182DM}.
Also the direct band gaps at $\Gamma$ of 1.70 eV in ZrS$_2$ and 2.08 eV in HfS$_2$ are perfectly in line with those hosted in Ref.~\cite{haas+182DM}. 
The two uppermost valence bands are degenerate at $\Gamma$ and approximately parabolic around it.
However, their different dispersion away from the zone center indicates the presence of charge carriers with different effective masses.
In both materials, the degeneracy of the VBM at $\Gamma$ is lifted upon inclusion of SOC, as shown in the insets of Figs.~\ref{fig:monobs}b) and~\ref{fig:monobs}d). 
The resulting splitting is 79 meV in ZrS$_2$ and 122 meV in HfS$_2$, respectively, also in agreement with results available in the literature~\cite{rasm-thyg15jpcc,haas+182DM,yagm+19prb}. 
Upon inclusion of SOC, the DFT gap, which remains indirect, is reduced by 40 meV in ZrS$_2$ and by 59 meV in HfS$_2$.
Also the lowest conduction band exhibits degeneracy at $\Gamma$, which is however not broken by SOC (see Figs.~\ref{fig:monobs}b and d).
The character of the electronic states in the vicinity of the gap is analyzed through the density of states (DOS), plotted in Fig.~\ref{fig:pdos}.
Therein, the total DOS is represented by the grey shaded area and the most relevant atom-projected contributions within the muffin-tin spheres by colored lines.
In both cases the valence band is dominated by the S $p$-states. 
In the conduction band the largest contributions come from the Zr and Hf $d$-states. 

A quantitative estimate of the band-gaps is obtained from $G_0W_0$ calculations carried out on top of PBE (Figs.~\ref{fig:monobs}a and c), resulting in 2.81 eV in ZrS$_2$ and in 2.62 eV in HfS$_2$.
In both cases, the fundamental gaps computed from $G_0W_0$ remain indirect, and their magnitude is almost twice as large as the PBE ones, corresponding to a QP correction of 1.6 eV for ZrS$_2$ and of 1.26 eV for HfS$_2$.
On the other hand, the self-energy correction to the direct gaps is systematically larger than the one to the fundamental gaps, \textit{i.e.}, by about 250-300 meV in both materials, resulting in a direct QP band gap of 3.52 eV in ZrS$_2$ and of 3.62 eV in HfS$_2$.
It is interesting to notice that both KS and QP direct gaps are larger in HfS$_2$ than in ZrS$_2$, while the indirect band gap in ZrS$_2$ is larger than the one in HfS$_2$.
The results reported in Ref.~\cite{haas+182DM} for both the fundamental and direct $G_0W_0$ gaps of the two monolayers show an analogous trend.
Moreover, as the $G_0W_0$ calculations in Ref.~\cite{haas+182DM} include SOC, it is possible for us to comment about this point.
In the case of ZrS$_2$, both the fundamental and the direct $G_0W_0$ gaps available in Ref.~\cite{haas+182DM} are larger than ours by a few tens of meV, thus suggesting that the influence of SOC is mild and analogous to the one discussed for PBE calculations.
On the other hand, the $G_0W_0$ gaps of HfS$_2$ in Ref.~\cite{haas+182DM} are 200 meV (direct gap) and 320 meV (fundamental gap) larger than our results.

\begin{figure*}
\includegraphics[width=\textwidth]{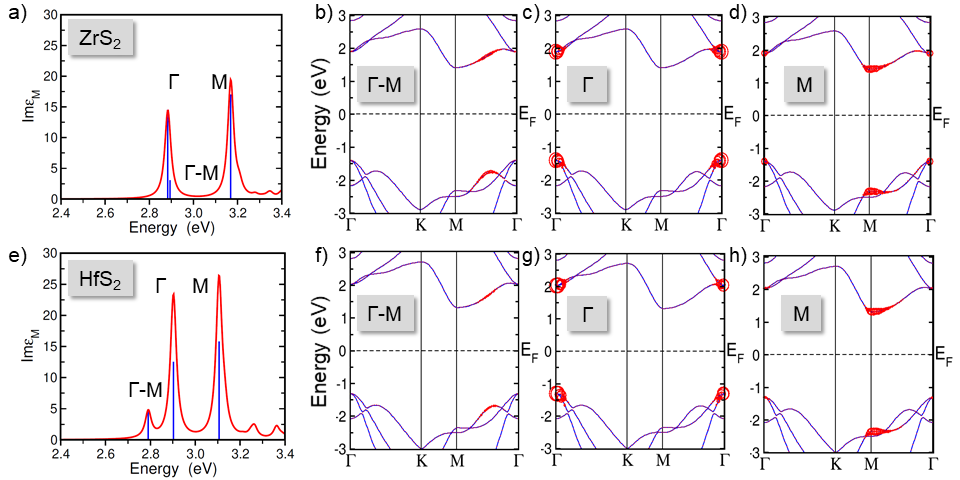}
\caption{BSE spectra of monolayer a) ZrS$_2$ and e) HfS$_2$, given by the in-plane component of the imaginary part of the dielectric function (Eq.~\ref{eq:ImeM}). Vertical bars denote the oscillator strengths of the main excitons. b) -- d) \textbf{k}-space analysis of the excitons forming the three main peaks in a). The same is shown for HfS$_2$ in panels f) -- h). The radius of the red circles is indicative of the relative weight of the electronic states contributing to the respective excitation.}
\label{fig:monobse}
\end{figure*}

The optical spectra of monolayer ZrS$_2$ and HfS$_2$ computed from the solution of the BSE are reported in Fig.~\ref{fig:monobse}. 
These calculations are performed including the QP particle correction to the direct band gap as a scissors operator.
The spectrum of ZrS$_2$ (upper panel) is dominated by two intense peaks formed by bound excitons with large oscillator strengths. 
These results are qualitatively different from those obtained within the independent-particle approximation in Ref.~\cite{vu+19sum}, pointing to the importance of an accurate treatment of many-body effects to correctly capture the optical properties of such systems.
The lowest-energy peak at 2.68 eV is formed by two bright excitations, marked by vertical lines in Fig.~\ref{fig:monobse}a). 
From the analysis of the \textbf{k}-space distribution of the exciton weights (Eqs.~\ref{eq:wh} and~\ref{eq:we}) shown in Figs.~\ref{fig:monobse}c) it is apparent that the first and more intense excitation originates from the VBM to the conduction band at $\Gamma$. 
When SOC is not included, this exciton is twofold degenerate because of the degeneracy of the VBM.
The second and weaker excitation forming the first peak stems from vertical transitions between points of the BZ that are along the $\Gamma$-M path. 
For this reason, we label this excitation $\Gamma$-M.
To avoid misunderstanding, we emphasize that it is computed at $\mathbf{q} \rightarrow 0$ like all the other excitations considered in this work. 
The second peak in the spectrum is given by a strong resonance which arises from transitions at the M point with non-negligible contributions also at $\Gamma$ (Fig. \ref{fig:monobse}d). 
Based on this analysis, we can determine the binding energies of these excitations as the difference between the excitation energy from the BSE and the energy difference between the involved QP states (see Table~\ref{table:EbX}). 
The binding energies of the two degenerate excitons at $\Gamma$ and of the exciton stemming from the transition between $\Gamma$ and M amount to 0.64 eV and 0.63 eV, respectively, while the exciton at M has a larger binding energy of 0.78 eV.

\begin{table}
\begin{ruledtabular}
\begin{tabular}{l c c r}
 &  & Exciton binding energies (eV)  \\
     & $\Gamma$ & $\Gamma$-M & M \\ \hline
 ZrS$_2$ & 0.64 & 0.63 & 0.78 \\ 
 HfS$_2$ & 0.73 & 0.84 & 0.86 \\ 
 ZrS$_2$/HfS$_2$ & 0.35 & 0.38 & 0.46
\end{tabular}
\end{ruledtabular}
\caption{Binding energies of the excitons marked in the spectra of ZrS$_2$, HfS$_2$, and ZrS$_2$/HfS$_2$, computed as the difference between the excitation energy from BSE and the QP energy of the corresponding vertical transitions.}
\label{table:EbX}
\end{table}

The spectrum of HfS$_2$ (Fig.~\ref{fig:monobse}e) is rather different from the one of ZrS$_2$. 
A pre-peak appears at approximately 2.5~eV, corresponding to an exciton stemming from transitions along the $\Gamma$-M path in the Brillouin zone (see Fig.~\ref{fig:monobse}f).
The excitations emerging from the vertical transitions at $\Gamma$ (also doubly degenerate) and M appear at higher energies, with larger oscillator strength that is comparable to their counterparts in the spectrum of ZrS$_2$.
However, different from ZrS$_2$, in HfS$_2$ the exciton labeled M in the spectrum of Fig.~\ref{fig:monobse}e) receives contributions only from transitions at M, with vanishing weights at $\Gamma$ (see Fig.~\ref{fig:monobse}h).
As shown in Table~\ref{table:EbX}, the binding energy of the exciton labeled M is the largest but almost comparable with the value obtained for the exciton distributed between $\Gamma$ and M. 
As a final remark, we emphasize that in Figs.~\ref{fig:monobse}a) and e) only the in-plane components of the imaginary part of the macroscopic dielectric function is reported. 
The out-of-plane counterpart (not shown) exhibits absorption maxima with over two orders magnitude smaller oscillator strengths.
Energetically, the absorption edges in the out-of-plane polarization direction are blue-shifted in both materials by 270 meV compared to the in-plane component.
This behavior, analogous to the one discussed recently for various prototypical monolayer systems~\cite{guil+19prb}, can be explained in terms of selection rules.
The electronic transitions contributing to the lowest-energy excitons (see Fig.~\ref{fig:monobse}) are dipole-allowed in the in-plane direction but forbidden in the out-of-plane one.
Hence, only higher-energy excitations actually contribute to the absorption in the normal direction.

\begin{figure}
\includegraphics[width=0.5\textwidth]{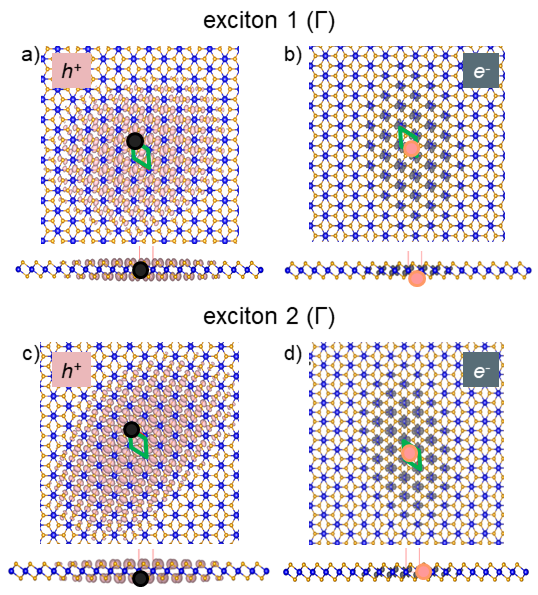}
\caption{Real-space distributions of the degenerate pair of excitons at the $\Gamma$ point in monolayer HfS$_2$ (top and side views shown). Hole (electron) distributions are shown in pink (grey). The position of the fixed coordinate is marked by a circle of the corresponding color. The unit cell is marked in green in the top views.}
\label{fig:realX}
\end{figure}

The exciton binding energies computed for the first excitations of HfS$_2$ are systematically larger compared to those of ZrS$_2$.
To understand this trend, we have to analyze the behavior of the dielectric function in the monolayers, which assumes the form $\varepsilon^{2D}(\mathbf{q}) = 1+ \alpha^{2D}\mathbf{q}$, where $\mathbf{q}$ are in-plane wave-vectors and $\alpha^{2D}$ is the in-plane static polarizability for the 2D material~\cite{keld79,cuda+11prb,berk+13prb,hues+13prb,cher+14prl,zhan+14prb}.
The values of $\alpha^{2D}$ in ZrS$_2$ and HfS$_2$ are 4.21 \AA{} and 3.60 \AA{}, respectively~\cite{haas+182DM}.
The larger value of $\alpha^{2D}$ in ZrS$_2$ compared to HfS$_2$ inserted in the expression of $\varepsilon^{2D}(\mathbf{q})$ above gives rise to an enhanced screening in ZrS$_2$ with respect to HfS$_2$, which in turn clarifies the observed trend for the binding energies in the two materials (see Table~\ref{table:EbX}).
All in all, the values reported in Table~\ref{table:EbX} for the lowest-energy excitation in both ZrS$_2$ and HfS$_2$ are in excellent agreement with the results in Ref.~\cite{haas+182DM}.

The analysis of the excitons in reciprocal space is complemented by the visualization of the e-h wave-functions (see Eq.~\ref{eq:exciton}). 
In Fig.~\ref{fig:realX} we plot  the square modulus of $\Psi^{\lambda}(\mathbf{r}_h, \mathbf{r}_e)$ for the twofold degenerate $\Gamma$ exciton in the spectrum of HfS$_2$.
The excitations in ZrS$_2$ exhibit an analogous behavior.
On the left (right) panels the plotted isosurface indicates the hole (electron) distribution for the fixed position of the electron (hole), represented by the colored dot.
The difference between the hole (Figs.~\ref{fig:realX}a and c) and electron distributions (Figs.~\ref{fig:realX}b and d) is quite remarkable in both cases.
The hole probability is extended over six unit cells in the in-plane direction with respect to the fixed electron position.
One of the excitons -- labeled exciton 1 ($\Gamma$) -- has an isotropic circular shape (Fig.~\ref{fig:realX}a), while the one -- labeled exciton 2 ($\Gamma$) -- a more elongated distribution (Fig.~\ref{fig:realX}c). 
This behavior can be rationalized considering the symmetry of the occupied $p$-orbitals that give rise to the hole distribution in the two degenerate excitons at $\Gamma$. 
The exciton plotted in Fig.~\ref{fig:realX}a) is given by transitions between bands with pronounced parabolic dispersion, while the exciton depicted in Fig.~\ref{fig:realX}c) stems from the bands with lower effective mass (see also Fig.~\ref{fig:monobse}).
The electron distributions shown in Figs.~\ref{fig:realX}b) and d) are considerably more localized compared to the ones of the hole shown in Figs.~\ref{fig:realX}a) and c).
In both cases, the electron probability extends over about four unit cells from the fixed hole position with a rather anisotropic shape that is compatible with the $d$-like character of the conduction states.
Since both of these degenerate excitons at $\Gamma$ correspond to transitions to the same unoccupied band, the character of the electron component is similar in the two cases. 

\subsection{ZrS$_2$/HfS$_2$ heterostructure}

\begin{figure}
\includegraphics[width=0.4\textwidth]{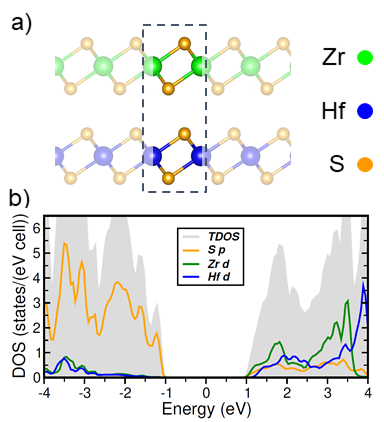}
\caption{a) Ball-and-stick representation of the ZrS$_2$/HfS$_2$ heterostructure in the AA stacking configuration, with blue, green and orange spheres denoting Zr, Hf, and S atoms respectively. b) Total and atom-projected QP density of states of the ZrS$_2$/HfS$_2$ heterostructure (only the most relevant contributions are shown).}
\label{fig:struct-pdos}
\end{figure}

The analysis of the electronic and optical properties of the ZrS$_2$ and HfS$_2$ monolayers is the starting point to characterize the heterostructure formed by these two materials.
Considering that the difference between the optimized lattice constants of ZrS$_2$ and HfS$_2$ is only 0.03~\AA{}, the heterobilayer is constructed assuming no lattice mismatch between the two monolayers~\cite{CoinLat}. 
By superimposing ZrS$_2$ and HfS$_2$, two trivial stacking patterns can be realized, usually indicated as AA and AB~\cite{ZrHfvdWhet}. 
In the AA stacking (see Fig.~\ref{fig:struct-pdos}a), the metal atoms are aligned vertically on top of each other, while in the AB stacking (not shown) the metal atoms of one layer are on top of the S atom of the other layer. 
We find that the AA stacking is energetically more favorable by 34 meV per unit cell compared to the AB variant, in good agreement with the result reported in Ref. \cite{ZrHfvdWhet}. 
In the following, we consider the ZrS$_2$/HfS$_2$ heterostructure only in the AA configuration, as depicted in Fig.~\ref{fig:struct-pdos}a). 
The in-plane lattice parameter is optimized and the residual interatomic forces are minimized.
The resulting lattice parameter $a$~=~3.66~\AA{} and the vertical separation of 5.80~\AA{} between the Hf and Zr atoms are slightly larger than the result obtained in Ref.~\cite{rehm-ding18pccp} with the PBE functional.

\begin{figure}
\includegraphics[width=0.4\textwidth]{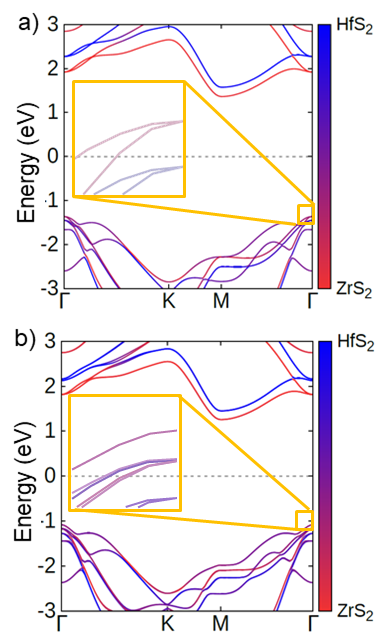}
\caption{Character-resolved band structures of the ZrS$_2$/HfS$_2$ heterostructure computed a) without SOC and b) with SOC. Degenerate a) and split-off bands b) at $\Gamma$ are highlighted in the insets. The QP correction is included through a scissors shift of 0.91 eV. }
\label{fig:hetbs}
\end{figure}

The electronic properties of the ZrS$_2$/HfS$_2$ heterostructure are analyzed in terms of character-resolved density of states (see Fig.~\ref{fig:struct-pdos}b) and band structure (Fig.~\ref{fig:hetbs}). 
In both cases, the QP correction to the fundamental gap is included as a scissors operator of 0.91 eV obtained from the $G_0W_0$ calculation.
The enhanced effective screening resulting from the stacking of the two monolayers reduces the size of the QP correction in the heterostructure by about 700 meV compared to ZrS$_2$ and by about 350 meV with respect to HfS$_2$.
Similar to the monolayers, also the heterostructure has an indirect band gap, with the CBm at M and the VBM at $\Gamma$, however, significantly reduced in size compared to isolated components.
The value resulting from our $G_0W_0$ calculation is 2.04~eV, in line with the one obtained in Ref.~\cite{rehm-ding18pccp} using the hybrid functional HSE06~\cite{HSE06}.
It can be ascribed to the enhanced screening that the two monolayers mutually exert on each other. 
The QP correction to the optical gap (1.24~eV) is larger than the QP correction to the fundamental gap (0.91 eV), like in the isolated monolayers -- see Section~\ref{mono}.
The direct gap from PBE is 1.68 eV while the one computed from $G_0W_0$ amounts to 2.92 eV.
From the DOS shown in Fig.~\ref{fig:struct-pdos}b) we notice that, similar to the isolated monolayers, the valence region is dominated by S $p$-contributions while the conduction band by the $d$-electrons of the metal atoms, yet with a small contribution from the S $p$-states. 
That result can be compared with the band structure computed without SOC shown in Fig.~\ref{fig:hetbs}a), where information about the band character is reported. 
In the adopted color code, red and blue lines denote pure ZrS$_2$ and HfS$_2$ contributions, respectively, while different shades of purple mark the hybridization between bands of the individual monolayer. 
The VBM has almost comparable contributions from both HfS$_2$ and ZrS$_2$ with the latter being slightly dominant. 
The degeneracy of the VBM at the $\Gamma$ point persists in spite of the hybridization. 
In the conduction region band hybridization is considerably reduced.
The lowest unoccupied band corresponds the first conduction band of ZrS$_2$ and the next one to the lowest conduction band of HfS$_2$.
Hence, the conduction band of the heterostructure is mainly a superposition of the conduction bands of ZrS$_2$ and HfS$_2$, with the former appearing at lower energy over almost the entire Brillouin zone.
Only approaching the zone center along the K-$\Gamma$ path these bands tend to hybridize (see Fig.~\ref{fig:hetbs}a). 
In the inset of Fig.~\ref{fig:hetbs}a), the degeneracy of the two parabolic valence bands at $\Gamma$ resembles the behavior discussed for the monolayers.
In the heterostructure, the additional electronic hybridization characterizing these bands gives rise to a degenerate pair of holes with different effective masses which are distributed across the two monolayers.

The inclusion of SOC does not alter the picture illustrated above (see Fig.~\ref{fig:hetbs}b). 
Similar to the monolayers, its main effect is to lift the degeneracy at the $\Gamma$ point, inducing a splitting of 83 meV, that effectively reduces the band gap by 53 meV. 
However, different from the monolayers, the effect of SOC is not limited to the VBM but affects also the next two lower valence bands at the $\Gamma$ point. 
When SOC is not considered, these two bands have more pronounced HfS$_2$ character and are degenerate at $\Gamma$ (see Fig.~\ref{fig:struct-pdos}b). 
With the inclusion of SOC (Fig.~\ref{fig:hetbs}b), their tendency to hybridize is enhanced. 
Albeit the VBM remains localized by more than 50\% on the ZrS$_2$ layer even upon inclusion of SOC, it is legitimate to consider the heterostructure still as type-I.
This result does not match the conclusion reported in Ref.~\cite{rehm-ding18pccp} that ZrS$_2$ and HfS$_2$ form a type-II heterostructure. 
That statement, however, was based on the analysis of the $d$-states, which are dominant at the bottom of the conduction band but are negligible compared to the sulfur $p$-states in the valence region (see Fig.~\ref{fig:struct-pdos}).
It is worth noting that the type-I character of the ZrS$_2$/HfS$_2$ heterostructure is consistent with the level alignment extrapolated from the values of the VBM and CBm with respect to the vacuum level in the isolated monolayers, as computed from HSE06 in Ref.~\cite{koda+18prb} and from $G_0W_0$ in Ref.~\cite{haas+182DM}.
In both ZrS$_2$ and HfS$_2$ the VBM is found at -7.50 eV, while the CBm of ZrS$_2$ is at -4.62 eV, which is 50 meV lower than the one of HfS$_2$.
This behavior is reproduced also by our results shown in Fig.~\ref{fig:hetbs}.

\begin{figure}
\includegraphics[width=0.5\textwidth]{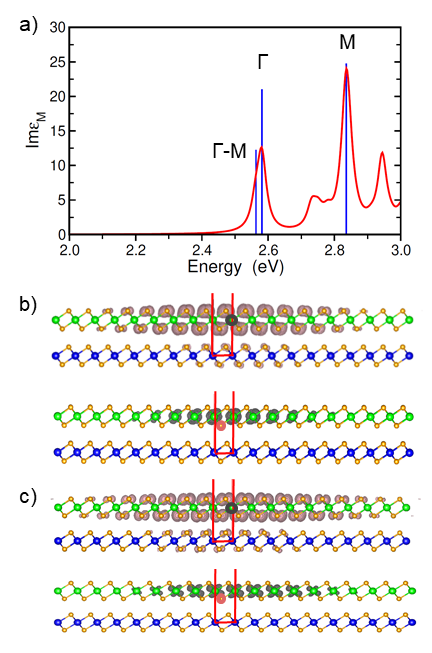}
\caption{a) BSE spectrum of the ZrS$_2$/HfS$_2$ heterostructure represented by the in-plane component of the imaginary part of the dielectric function with the most relevant excitonic peaks labeled according to their character. b) and c) Real-space distribution of the two degenerate excitons at the $\Gamma$ point: upper (lower) panels depict the hole (electron) distribution with the fixed electron (hole) position indicated by the circles. The color code for atoms and isosurfaces is the same as in Fig.~\ref{fig:struct-pdos}a) and Fig.~\ref{fig:realX}, respectively.}
\label{fig:hetbse}
\end{figure}

The optical spectrum of the ZrS$_2$/HfS$_2$ heterostructure, shown in Fig.~\ref{fig:hetbse}a), is evidently not a superposition of the spectra of the two isolated monolayers.
Some important differences compared to the results in Figs.~\ref{fig:monobse}a) and e) are noticeable.
First, in the heterostructure the absorption onset is red-shifted by about 200 meV, consistent with the reduction of the QP gap. 
Second, the first peak appearing in Fig.~\ref{fig:hetbse}a) is formed by two bright excitons: The energetically lower-lying one stems from transitions between $\Gamma$ and M, while the second one is twofold degenerate and arises from transitions around $\Gamma$.
Also in this case the out-of-plane component of the macroscopic dielectric function (not shown) is blue-shifted compared to the in-plane one.
However, due to the enhanced screening in the heterostructure, such blue-shift amounts to only 150 meV.
This corresponds to a reduction of 120 meV compared to the isolated monolayers.
Above the first peak, we find a weaker resonance at about 2.4 eV, which is followed by an intense maximum at 2.5 eV, corresponding to transitions at the M point. 
The binding energies of all these excitations, reported in the third row of Table~\ref{table:EbX}, are significantly smaller than their counterparts in the monolayers, consistent with the increased screening in the heterostructure compared to the isolated monolayers.

By plotting the exciton wavefunction (see Eq.~\ref{eq:exciton}) with fixed particle and hole position, respectively, it is possible to analyze the real-space distribution of the correlated hole and electron probabilities associated to the pair of degenerate excitons at the $\Gamma$ point.
In this way, one can visualize the (de)localization of the photo-excited charge carriers that are formed when the heterostructure is impinged by light. 
Consistent with the predominant type-I level alignment discussed above, both the hole and the electron distributions are mainly localized on the ZrS$_2$ layer.
On the one hand, the hole contributions to the e-h pairs are extended over approximately 10 unit cells in the plane and retain the shape of the S $p$-orbitals.
On the other hand, the electron contributions are more localized around the (fixed) position of the hole, extending about 7 unit cells around it, similar to the localized interlayer excitons identified in bulk h-BN~\cite{aggo+18prb}. 
Moreover, while the electron is solely distributed on the ZrS$_2$ layer, the hole exhibits non-negligible probability also on the HfS$_2$ sheet.
These findings reveal the non-trivial behavior of the photo-excited charge carriers in the ZrS$_2$/HfS$_2$ heterostructure.

\section{Summary and Conclusions}
We have studied the electronic and optical properties of monolayer ZrS$_2$ and HfS$_2$ and their vdW heterostructure by means of DFT and MBPT, discussing the role of SOC and excitonic effects.
The isolated monolayers are indirect semiconductors with QP gaps of 2.81 eV (ZrS$_2$) and 2.62 eV (HfS$_2$).
SOC leads to the splitting of the uppermost valence band leaving the lowest conduction band unaltered.
The optical spectra of these systems exhibit pronounced excitonic peaks with different characteristics of the specific material.
While the first exciton in the spectrum of ZrS$_2$ is given by a transition at the $\Gamma$ point, in HfS$_2$ the first bright peak stems from vertical transitions between $\Gamma$ and M.
The indirect nature of the band-gap in the monolayers is retained also in the heterostructure, where the valence bands are strongly hybridized while the two lowest conduction bands have almost pure ZrS$_2$ and HfS$_2$ character, respectively.
These characteristics make the ZrS$_2$/HfS$_2$ bilayer a type-I heterostructure, with the lowest-energy electron-hole pairs being mostly localized on the ZrS$_2$ layer.
However, the non-negligible probability to find the hole also in HfS$_2$ points to a non-trivial behavior of the photo-excited charge carriers in this system.
The exciton binding energy of the lowest-lying exciton, which amounts to 0.36 eV, is significantly reduced compared to the monolayers, suggesting that the electron-hole pairs can be more easily dissociated in the heterostructure compared to the isolated ZrS$_2$ and HfS$_2$ sheets. 

These findings have relevant implications in view of the opto-electronic applications of this novel class of low-dimensional materials. 
In light of the recent reports on prototypical devices based on group-IV TMDCs~\cite{ZrS2sol,4TMDrev,EnSciZrS2het,HfSe2FET_1,HfS2FET_1,HfS2FET_2}, our results suggest that by replacing the ZrS$_2$ component with the ZrS$_2$/HfS$_2$ heterostructure, dissociation of the intra-layer excitons within ZrS$_2$ can be significantly enhanced, leading to a higher photoelectron generation rate. 
The quantitative analysis of the electronic and optical properties of group-IV TMDCs and their vdW heterostructure reported in this paper contributes to further understand, predict, and tailor the photo-physical behavior of these promising materials for the next generation of opto-electronic devices.

\section*{Acknowledgment}
Work supported by the Deutscher Akademischer Austauschdienst (DAAD) and the International Max Planck Research School (IMPRS) on Functional Interfaces in Physics and Chemistry. CD and CC appreciate partial funding from the Deutsche Forschungsgemeinschaft (DFG) - Projektnummer 182087777 - SFB 951.

\bibliographystyle{apsrev4-1}

%

\end{document}